\documentclass[runningheads]{llncs}
\usepackage{caption}
\usepackage{multirow}
\usepackage{graphicx}
\usepackage{amsmath}
\usepackage{bm}
\usepackage{subfigure}
\begin{document}

\title{Explore Entity Embedding Effectiveness in Entity Retrieval}

\author{Zhenghao Liu\inst{1} \and
Chenyan Xiong\inst{2} \and
Maosong Sun\inst{1}\thanks{Corresponding author: M. Sun (sms@tsinghua.edu.cn)} \and
Zhiyuan Liu\inst{1}}

\authorrunning{Z. Liu et al.}

\institute{Department of Computer Science and Technology\\
Institute for Artificial Intelligence\\
State Key Lab on Intelligent Technology and Systems \\
Tsinghua University, Beijing, China \\
 \and
Microsoft Research AI, Redmond, USA.}

\maketitle              
\begin{abstract}
This paper explores entity embedding effectiveness in ad-hoc entity retrieval, which introduces distributed representation of entities into entity retrieval. The knowledge graph contains lots of knowledge and models entity semantic relations with the well-formed structural representation. Entity embedding learns lots of semantic information from the knowledge graph and represents entities with a low-dimensional representation, which provides an opportunity to establish interactions between query related entities and candidate entities for entity retrieval. Our experiments demonstrate the effectiveness of entity embedding based model, which achieves more than 5\% improvement than the previous state-of-the-art learning to rank based entity retrieval model. Our further analysis reveals that the entity semantic match feature effective, especially for the scenario which needs more semantic understanding.

\end{abstract}

\section{Introduction}
In the past decade, large-scale public knowledge bases have emerged, such as DBpedia~\cite{balog2013test}, Freebase~\cite{balog2011query} and Wikidata~\cite{bendersky2010learning}. These knowledge bases provide a well-structured knowledge representation and have become one of the most popular resources for many applications, such as web search and question answering. A fundamental process in these systems is ad-hoc entity retrieval, which has encouraged the development of entity retrieval systems. Ad-hoc entity retrieval in the web of data (ERWD) aims to answer user queries through returning entities from publicly available knowledge bases and satisfy some underlying information need.

Knowledge bases represent knowledge with Resource Description Framework (RDF) triples for structural information. Entity related triples contain lots of related information, such as name, alias, category, description and relationship with other entities. Previous entity retrieval works represent an entity by grouping entity related triples into different categories. And the multi-field entity representation provides an opportunity to convert the entity retrieval task to a document retrieval task. Therefore, lots of ranking methods can be leveraged, such as BM25, TF-IDF and Sequential Dependence Models (SDM). Learning to rank (LeToR) models provide an effective way to incorporate different match features and achieve the state-of-the-art for entity retrieval~\cite{chen2016empirical}. These entity retrieval systems only leverage text based matches and neglect entity semantics in the knowledge graph. Therefore, field representation shows its limitation with the structural knowledge representation.

Knowledge representation learning provides an effective way to model entity relations with embedding.  The relations of entities in a knowledge graph are stored in RDF triples which consist of the head entity, relation and tail entity. Previous works, such as TransE~\cite{bordes2013translating}, represent both entities and relations as the low-dimensional representation. Then they formalize the entities and relations with different energy functions. Knowledge representation learning helps to learn the structural information of the knowledge graph, which can better help entity retrieval models understand the semantic information from entities.

This work investigates the effectiveness of entity embedding, which contains knowledge graph semantics, for entity retrieval. It utilizes TransE to get the low-dimensional representation for each entity. And then we calculate the soft match feature between query entities and candidate entities. Furthermore, we also follow the previous methods to represent entities textual information with multiple fields and exact match features with different ranking methods for all fields. The learning to rank models is utilized to combine all exact match features and entity soft match feature for the ranking score. Experiments on an entity search test benchmark confirm that entity embedding based soft match feature is critical for entity retrieval and significantly improve the previous state-of-the-art entity retrieval methods by over 5\%. Our analyses also indicate that entity embedding based semantic match plays an important role, especially for the scenario which needs more linguistic and semantic understanding. We released all resources of data and codes via github\footnote{\url{https://github.com/thunlp/EmbeddingEntityRetrieval}}.

\section{Related Work}
The ERWD task, which is first introduced by Pound et al.~\cite{pound2010ad}, focuses on how to answer arbitrary keyword queries by finding one or more entities with entity representations. Existing entity retrieval systems concern more about the representation of entities. Early works, especially in the context of expert search, obtain entity representations by considering mentions of the given entity~\cite{balog2006formal,balog2011query}. The INEX 2007-2009 Entity Retrieval track (INEX-XER)~\cite{de2007overview,demartini2009overview} studies entity retrieval in Wikipedia, while the INEX 2012 Linked Data track further considers Wikipedia articles together with RDF properties from the DBpedia and YAGO2 knowledge bases~\cite{wang2012overview}. The recent works usually represent entities as the fielded document~\cite{balog2013test,zhiltsov2015fielded}, which divides the entity representation into three or five categories. These entity representation methods provide a possible way to solve the entity retrieval problem with document retrieval methods.

Previous document retrieval models calculate the query and document relevance with bag-of-word representations, such as BM25 and Language Model (LM). Nevertheless, these bag-of-words retrieval models neglect term dependence, which is an important match signal for document retrieval. Markov Random Field (MRF)~\cite{metzler2005markov} for document retrieval provides a solid theoretical way to model the dependence among query terms. Sequential Dependence Model (SDM)~\cite{metzler2005markov} is a variation of Markov Random Field, which considers unigram, ordered bigram and unordered bigram match features. The SDM provides a good balance between retrieval effectiveness and efficiency.

Entity retrieval models leverage document retrieval models and extend them to multiple fields. They weight all ranking scores from all categories of the entity representation for the ranking score. They mainly leverage the standard bag-of-words framework to calculate the similarity between query and candidate entities. BM25F~\cite{robertson2004simple} and Mixture of Language Models (MLM)~\cite{ogilvie2003combining} combine BM25 and Language Model retrieval models to the multi-field entity retrieval. Different from MLM and BM25F, Fielded Sequential Dependence Model (FSDM)~\cite{zhiltsov2015fielded} considers the sequential dependence and leverages SDM to calculate the relevant score between query and each field of the candidate entity. On the other hand, Probabilistic Retrieval Model for Semistructured Data (PRMS)~\cite{kim2009probabilistic} weights query terms according to document collection statistics for the better retrieval performance.  To further leverage the entity based interactions between the query and candidate entities, some works~\cite{hasibi2016exploiting} calculate the entity mention based exact match feature between query and candidate entities. State-of-the-art learning to rank models, such as Coordinate Ascent and RankSVM, provide an opportunity to combine features from different models and different fields, which achieves the state-of-the-art for entity retrieval~\cite{chen2016empirical}.  

The knowledge representation learning methods model entity structural information and encode entities into a low-dimensional vector space. TransE~\cite{bordes2013translating} is one of the most popular and robust works for knowledge representation learning. TransE interprets knowledge graph triples as a translation: the entity vector plus the relation vector is equal to the tail entity. Moreover, the entity embedding with semantic information has further improved ranking performance~\cite{liu2018entity} for ad-hoc retrieval. Therefore, knowledge embedding may provide a potential way to bring entity semantic information from knowledge graph to entity retrieval.

\section{Methodology}
% overall introduction.
In this section, we introduce the text match based retrieval model, entity mention based retrieval model and our entity embedding based model. Given a query $Q = \{q_1, q_2,..., q_n\}$ and an entity representation $E$, our aim is to generate a ranking score $f(Q, E)$ to rank candidate entities.

\subsection{Text based Retrieval Model}
Existing entity search models leverage term and term dependence based match features to calculate $Q$ and $E$ similarity based on the Markov Random Field (MRF) model. Therefore, we introduce two variations of MRF, the Sequential Dependence Model (SDM) and Fielded Sequential Dependence Model (FSDM) in this part.

\subsubsection{Sequential Dependence Model.}
The Sequential Dependence Model (SDM) considers both unigram and bigram match features for ranking. To calculate the ranking score, it is apparent that computing the following posterior probability is sufficient:
\begin{equation}
P(E|Q) = \frac{P(Q, E)}{P(Q)}\overset{rank}{=} P(Q, E).
\end{equation}

Based on MRF model, we could get query term and adjacent term cliques. Then we incorporate the term $q_i$ based match feature and the term dependence based match feature to get the SDM ranking function:
\begin{equation}
     P(E|Q) \overset{rank}{=}\lambda_T\sum_{q_i\in Q}f_T(q_i,E)+\lambda_O\sum_{q_i,q_{i+1}\in Q}f_O(q_i, q_{i+1},E)+\lambda_U\sum_{q_i,q_{i+1}\in Q}f_U(q_i, q_{i+1},E),
\end{equation}
where $\lambda$ is the parameter to weight features, which should meet $\lambda_T+\lambda_O+\lambda_U = 1$. $f_T(q_i,E)$ denotes unigram match feature. $f_O(q_i, q_{i+1},E)$ and $f_U(q_i, q_{i+1},E)$ represent ordered and unordered bigram match features respectively. Then the specific feature functions are presented as follow: 
\begin{align}
f_T(q_i, E) &= \log [\frac{{tf}_{q_i, E}+ \mu \frac{{cf}_{q_i}}{|C|}}{|E|+\mu}],\\
f_O(q_i, q_{i+1}, E) &= \log [\frac{{tf}_{\#1(q_i,q_{i+1}),E}+ \mu \frac{{cf}_{\#1(q_i,q_{i+1})}}{|C|}}{|E|+\mu}],\\
f_U(q_i, q_{i+1}, E) &= \log [\frac{{tf}_{\#uwN(q_i,q_{i+1}),E}+ \mu \frac{{cf}_{\#uwN(q_i,q_{i+1})}}{|C|}}{|E|+\mu}],
\end{align}
where $tf$ and $cf$ denotes uni-gram or bi-gram term frequency for the candidate entity and entire entity collection respectively. $(\#1(q_i,q_{i+1}),E)$ calculates the exact match for $q_i, q_{i+1}$ and $(\#uwN(q_i,q_{i+1}),E)$ counts the number of co-occurrence times of $q_i$ and $q_{i+1}$ within a $N$ size window. And $\mu$ is the Dirichlet prior.

\subsubsection{Field Sequential Dependence Model.}
Field Sequential Dependence Model extends the SDM with a Mixture of Language Model (MLM) for each field of an entity representation. MLM computes each field probability and combines all fields with a linear function. For the field $f \in F$, the match feature functions $f_T(q_i,E)$, $f_O(q_i, q_{i+1},E)$ and $f_U(q_i, q_{i+1},E)$ can be extended as follow:

\begin{align}
f_T(q_i, E) &= \log\sum_{f}w_f^T \frac{{tf}_{q_i, E_f}+ \mu_f \frac{{cf}_{q_i,f}}{|C_f|}}{|E_f|+\mu_f},\\
f_O(q_i, q_{i+1}, E) &= \log \sum_{f}w_f^O\frac{{tf}_{\#1(q_i,q_{i+1}),E_f}+ \mu_f \frac{{cf}_{\#1(q_i,q_{i+1}),f}}{|C_f|}}{|E_f|+\mu_f},\\
f_U(q_i, q_{i+1}, E) &= \log\sum_{f}w_f^U\frac{{tf}_{\#uwN(q_i,q_{i+1}),E_f}+ \mu_f \frac{{cf}_{\#uwN(q_i,q_{i+1}),f}}{|C_f|}}{|E_f|+\mu_f},
\end{align}
where $\mu_f$ is the weight for the field $f$. The FSDM represents entities with a novel five-field schema: The \texttt{names} contains entity names, such as the label relation from RDF triples; The \texttt{attributes} field involves all text information, such as entity abstract, except entity \texttt{names} field; The \texttt{categories} field implies entity categories; Then the \texttt{SimEn} and \texttt{RelEn} denote similar or aggregated entity and related entity respectively. Then FSDM weights all field weight and achieves a further improvement than SDM. 

\subsection{Entity Mention based Retrieval Model}
In this part, we introduce the entity mention based retrieval model for entity retrieval. The Entity Linking incorporated Retrieval (ELR) model represents the query $Q$ through an annotated entity set $\hat{E}(Q)=\{e_1, e_2, ..., e_m\}$ with the confidence score $s(e_i)$ for each entity. According to MRF graph, ELR involves interactions between $\hat{E}(Q)$ and $E$. Then ELR extends MRF with a linear combination of correlate entity potential function:
\begin{equation}
    \begin{split}
        P(E|Q) \overset{rank}{=}& \sum_{q_i\in Q}\lambda_T f_T(q_i,E)+\sum_{q_i,q_{i+1}\in Q}\lambda_O f_O(q_i, q_{i+1},E)+\\
&\sum_{q_i,q_{i+1}\in Q}\lambda_U f_U(q_i, q_{i+1},E)+\sum_{e\in \hat{E}(Q)}\lambda_{\hat{E}} f_{\hat{E}}(e, E),
    \end{split}
\end{equation}
Then ELR takes the entity confidence score to weight all entity based matches:
\begin{equation}
    \begin{split}
P(E|Q) \overset{rank}{=}& \lambda_T\sum_{q_i\in Q} \frac{1}{|Q|}f_T(q_i,E)+\lambda_O \sum_{q_i,q_{i+1}\in Q}\frac{1}{|Q|-1} f_O(q_i, q_{i+1},E)+\\
&\lambda_U \sum_{q_i,q_{i+1}\in Q}\frac{1}{|Q|-1} f_U(q_i, q_{i+1},E)+\lambda_E \sum_{e\in \hat{E}(Q)}s(e)f_{\hat{E}}(e,E),
    \end{split}
\end{equation}
where $|Q|$ and $|Q|-1$ are utilized to smooth TF features according to the sequence length. 
% Figure.
% Table features.
\subsection{Entity Embedding based Retrieval Model}
Previous entity retrieval models only calculate $Q$ and $E$ correlation with exact matches without considering knowledge based semantic information. For example, given two entities \texttt{Ann Dunham} and \texttt{Barack Obama}, \texttt{Ann Dunham} is a parent of \texttt{Barack Obama}. It is inevitable that exact matches will regard \texttt{Ann Dunham} and \texttt{Barack Obama} as different entities. To solve this problem, we leverage entity embeddings with knowledge graph semantics to improve entity retrieval performance.

To leverage knowledge embedding to calculate our ranking features, we first map both entities in $\hat{E}(Q)$ and $E$ into the same vector space. Then we also consider confidence score $s(e_i)$ for the similarity of $i$-th entity in the query annotated entity set $\hat{E}(Q)$ and the candidate entity $E$:
\begin{equation}
f(Q, E) = \sum_{i=1}^{m} s(e_i)\cdot\cos(\vec{v}_{e_i}, \vec{v}_E),
\end{equation}
where $\vec{v}_{e_i}$ and $\vec{v}_E$ TansE embedding for entity $e_i$ and $E$ respectively. 

Translation based methods successfully model structural knowledge bases for entity representation learning. TransE is a robust and efficient algorithm in translation-based entity embedding model and we use TransE to obtain the entity embedding. For the entity triple in knowledge base $S$, the tail entity $t$ should be close to the head entity $h$ plus the relationship $r$. The energy function is demonstrated as follows:
\begin{equation}
J(h,r,t) = ||h + r - t||_+.
\end{equation}
Then we minimize the pairwise loss function over the training set to optimize both entity and relation embeddings:
\begin{equation}
L = \sum \limits_{(h, r, t) \in S} \sum \limits_{(h^{'}, r, t^{'}) \in \hat{S}} [\gamma + d(\vec{v}_h + \vec{v}_r, \vec{v}_t) - d(\vec{v}_{h^{'}} + \vec{v}_{r}, \vec{v}_{{t}^{'}})],
\end{equation}
where $h^{'}$ and $t^{'}$ denote negative head entity and tail entity for the relation $r$. $d$ denotes the distance between two vectors.

\section{Experimental Methodology}
This section describes our experimental methods and materials, including dataset, baselines and parameters setting. 
\subsection{Dataset}
We use DBpedia version 3.7 as our knowledge base and compare the effectiveness of our knowledge embedding model based on a publicly available benchmark which contains 485 queries and 13090 related entities shown as Table \ref{table2}. There are four types of queries in this collection: \texttt{Entity} (e.g. ``NAACP Image Awards''), \texttt{Type} (e.g. ``circus mammals''), \texttt{Attribute} (e.g. ``country German language'') and \texttt{Relation} (e.g. ``Which airports are located in California, USA''). Therefore, the four subtasks for entity retrieval evaluate models from different aspect:

$\bullet$ \textbf{SemSearch ES:} Queries usually consist of named entity. And queries are oriented to the specific entities, which usually need to be disambiguated. (e.g., ``harry potter'', ``harry potter movie'')

$\bullet$ \textbf{ListSearch:} A query set combines INEX-XER, SemSearch LS, TREC Entity queries. This subtask aims to a list of entities that matches a certain criteria. (e.g. ``Airlines that currently use Boeing 747 planes'')

$\bullet$ \textbf{INEX-LD:} IR-style keyword queries, including a mixture of \texttt{Entity}, \texttt{Type}, \texttt{Attribute} and \texttt{Relation}. (e.g., ``bicycle sport races'')

$\bullet$ \textbf{QALD-2:} Consisting of natural language questions as well as involve four different types. (e.g., ``Who wrote the book The pillars of the Earth'')

The SemSearch ES, ListSearch and QALD-2 subtasks need more linguistic or semantic understanding. On the other hand, INEX-LD focuses more on keyword matches.
\begin{table}
\centering
\caption{Statistic of DBpedia-entity test collection}\label{table2}
\setlength{\tabcolsep}{1.9mm}{
\begin{tabular}{llll}
\hline
\textbf{Query set}&\textbf{\#queries}&\textbf{\#rel}&\textbf{Query types}\\
\hline
SemSearch ES&130&1131&Entity\\
ListSearch&115&2398&Type\\
INEX\_LD&100&3756&Entity, Type, Attribute, Relation\\
QALD-2&140&5805&Entity, Type, Attribute, Relation\\
\hline
\textbf{Total}&485&13090&-\\
\hline
\end{tabular}}
\end{table}

\begin{table}
\caption{Traditional baseline features.}\label{table1}
\centering
\begin{tabular}{ll}
\hline
\textbf{Features}&\textbf{Dimension}\\
\hline
FSDM&1\\
\hline
SDM on all fields&5\\
\hline
BM25 on all fields&5\\
\hline
Language model on all fields&5\\
\hline
Coordinate match on all fields&5\\
\hline
Cosine similarity on all fields&5\\
\hline
\end{tabular}
\end{table}

\subsection{Baselines}
We follow the previous state-of-the-art model~\cite{chen2016empirical} and leverage feature based learning to rank methods, RankSVM and Coordinate Ascent, as our ranking algorithms. Our baseline methods also utilize 26 features from different traditional ranking models and different fields, as shown in Table \ref{table1}. 

The entity mention match feature is also incorporated with word based 26 ranking features for our baseline, denoted as ``\texttt{+ELR}''. The entity mention based match feature only considers exact matches for entity mention. And we further incorporate our entity embedding based semantic match feature with the baseline 26 features and is denoted as ``\texttt{+TransE}''.

\subsection{Implementation Details}
We use the Fielded Sequential Dependence Model (FSDM) as the basic retrieval model to generate the candidate entity set with top 100 entities. And all models in our experiments rerank candidate entities. RankSVM implementation is provided
by SVMLight toolkit\footnote{\url{https://www.cs.cornell.edu/people/tj/svm\_light/svm\_rank.html}}. Coordinate Ascent implementation is provided by RankLib\footnote{\url{http://sourceforge.net/p/lemur/wiki/RankLib/}}. All models in our experiments are trained and tested using five fold cross validation and all models keep the same partition. Moreover, all parameter settings are kept the same with the previous work~\cite{chen2016empirical}. All methods are evaluated by MAP@100, P@10, and P@20. Statistic significances are tested by permutation test with P$<0.05$.

For entity embedding, we involve 11,988,202 entities to train our TransE model. The TransE model is implemented with C++ language\footnote{\url{https://github.com/thunlp/Fast-TransX}}.  We set the entity dimension as 100 dimension. All embeddings are optimized with SGD optimizer and 0.001 learning rate.

\section{Evaluation Result}
In this section, we present the model performance and the feature weight distribution to demonstrate the effectiveness of our model.
\begin{table}[t]
\centering
\caption{Entity retrieval performance with Coordinate Ascent. Relative performances compared are in percentages. $\dagger$, $\ddagger$ indicate statistically significant improvements over Baseline$^{\dagger}$ and +ELR$^{\ddagger}$ respectively.}\label{table:ca}
\resizebox{\textwidth}{!}{\setlength{\tabcolsep}{1.1mm}{
\begin{tabular}{l|l r|l r|l r| c}
\hline
\multirow{2}{*}{ \textbf{models}} & \multicolumn{7}{c}{ \textbf{SemSearch ES}}\\
\cline{2-8}
& \multicolumn{2}{c|}{\textbf{MAP}} & \multicolumn{2}{c|}{\textbf{P@10}} & \multicolumn{2}{c|}{\textbf{P@20}}&\textbf{W/T/L}\\
\hline
Baseline& $0.3899$ & -- & $0.2908$ & -- & $0.2077$ & -- & --/--/--\\
+ELR & ${0.3880}$ & $ -0.49\%  $ & \bm{${0.3023}^{\dagger }$ }& $ +3.95\%  $ & ${0.2150}^{\dagger }$ & $ +3.51\%  $ & 58/19/53\\
+TransE& \bm{${0.4085}^{\ddagger }$} & $ +4.77\%  $ & \bm{${0.3023}^{\dagger }$} & $ +3.95\%  $ & \bm{${0.2165}^{\dagger }$} & $ +4.24\%  $ & 64/16/50\\
\hline
\multirow{2}{*}{} & \multicolumn{7}{c}{ \textbf{ListSearch}}\\

\cline{2-8}
& \multicolumn{2}{c|}{\textbf{MAP}} & \multicolumn{2}{c|}{\textbf{P@10}} & \multicolumn{2}{c|}{\textbf{P@20}}&\textbf{W/T/L}\\
\hline
Baseline & $0.2334$ & -- & $0.3130$ & -- & $0.2378$ & -- & --/--/--\\
+ELR & ${0.2443}^{\dagger }$ & $ +4.67\%  $ & ${0.3130}$ & $ +0.00\%  $ & ${0.2422}$ & $ +1.85\%  $ & 54/22/39\\
+TransE & \bm{${0.2507}^{\dagger }$} & $ +7.41\%  $ & \bm{${0.3304}^{\dagger }$} & $ +5.56\%  $ & \bm{${0.2543}^{\dagger \ddagger }$} & $ +6.94\%  $ & 65/20/30\\
\hline
\multirow{2}{*}{} & \multicolumn{7}{c}{ \textbf{INDEX-LD}}\\
\cline{2-8}
& \multicolumn{2}{c|}{\textbf{MAP}} & \multicolumn{2}{c|}{\textbf{P@10}} & \multicolumn{2}{c|}{\textbf{P@20}}&\textbf{W/T/L}\\
\hline
Baseline & $0.1298$ & -- & $0.2900$ & -- & $0.2285$ & -- & --/--/--\\
+ELR & ${0.1275}$ & $ -1.77\%  $ & \bm{${0.2920}$} & $ +0.69\%  $ & \bm{${0.2335}$} & $ +2.19\%  $ & 44/10/46\\
+TransE & \bm{${0.1312}$} & $ +1.08\%  $ & ${0.2860}$ & $ -1.38\%  $ & ${0.2255}$ & $ -1.31\%  $ & 42/12/46\\
\hline
\multirow{2}{*}{} & \multicolumn{7}{c}{ \textbf{QALD-2}}\\
\cline{2-8}
& \multicolumn{2}{c|}{\textbf{MAP}} & \multicolumn{2}{c|}{\textbf{P@10}} & \multicolumn{2}{c|}{\textbf{P@20}}&\textbf{W/T/L}\\
\hline
Baseline& $0.1998$ & -- & $0.1500$ & -- & $0.1196$ & --& --/--/--\\
+ELR& ${0.2074}$ & $ +3.80\%  $ & ${0.1664}$ & $ +10.93\%  $ & ${0.1282}$ & $ +7.19\%  $ & 45/59/36\\
+TransE& \bm{${0.2270}^{\dagger }$} & $ +13.61\%  $ & \bm{${0.1700}^{\dagger }$} & $ +13.33\%  $ & \bm{${0.1371}^{\dagger \ddagger }$} & $ +14.63\%  $ & 48/62/30\\
\hline
\multirow{2}{*}{} & \multicolumn{7}{c}{ \textbf{ALL}}\\

\cline{2-8}
& \multicolumn{2}{c|}{\textbf{MAP}} & \multicolumn{2}{c|}{\textbf{P@10}} & \multicolumn{2}{c|}{\textbf{P@20}}&\textbf{W/T/L}\\
\hline
Baseline & $0.2454$ & --& $0.2540$ & -- & $0.1934$ & -- & --/--/--\\
+ELR& ${0.2472}$ & $ +0.73\%  $& ${0.2544}$ & $ +0.16\%  $& ${0.1945}$ & $ +0.57\%  $& 175/145/165\\
+TransE& \bm{${0.2597}^{\dagger \ddagger }$} & $ +5.83\%  $& \bm{${0.2639}^{\dagger \ddagger }$} & $ +3.90\%  $& \bm{${0.1970}^{\dagger }$} & $ +1.86\%  $& 212/122/151\\
\hline
\end{tabular}}}
\end{table}

\begin{table}[t]
\caption{Entity retrieval performance with RankSVM. Relative performances compared are in percentages. $\dagger$, $\ddagger$ indicate statistically significant improvements over Baseline$^{\dagger}$ and +ELR$^{\ddagger}$ respectively.}\label{table:ranksvm}
\setlength{\tabcolsep}{1.2mm}{
\begin{tabular}{l|l r|l r|l r | c}

\hline
\multirow{2}{*}{ \textbf{models}} & \multicolumn{7}{c}{ \textbf{SemSearch ES}}\\
\cline{2-8}
& \multicolumn{2}{c|}{\textbf{MAP}} & \multicolumn{2}{c|}{\textbf{P@10}} & \multicolumn{2}{c|}{\textbf{P@20}}&\textbf{W/T/L}\\
\hline
Baseline & $0.3895$ & -- & $0.3038$ & -- & $0.2169$ & -- & --/--/--\\
+ELR & ${0.3881}$ & $ -0.36\%  $ & ${0.3046}$ & $ +0.26\%  $ & ${0.2173}$ & $ +0.18\%  $ & 48/38/44\\
+TransE & \bm{${0.4061}^{\dagger \ddagger }$} & $ +4.26\% $ & \bm{${0.3077}$} & $ +1.28\%  $& \bm{${0.2196}$} & $ +1.24\%  $ & 52/24/54\\
\hline
\multirow{2}{*}{} & \multicolumn{7}{c}{ \textbf{ListSearch}}\\
\cline{2-8}
& \multicolumn{2}{c|}{\textbf{MAP}} & \multicolumn{2}{c|}{\textbf{P@10}} & \multicolumn{2}{c|}{\textbf{P@20}}&\textbf{W/T/L} \\
\hline
Baseline& $0.2323$ & --  & $0.3078$ & -- & $0.2513$ & -- & --/--/--\\
+ELR & ${0.2390}^{\dagger }$ & $ +2.88\%  $ & ${0.3148}$ & $ +2.27\%  $ & ${0.2530}$ & $ +0.68\%  $ & 60/25/30\\
+TransE & \bm{${0.2439}^{\dagger }$} & $ +4.99\%  $  & \bm{${0.3252}^{\dagger }$} & $ +5.65\%  $ & \bm{${0.2565}$} & $ +2.07\%  $ & 54/23/38\\
\hline
\multirow{2}{*}{} & \multicolumn{7}{c}{ \textbf{INDEX-LD}}\\
\cline{2-8}
& \multicolumn{2}{c|}{\textbf{MAP}} & \multicolumn{2}{c|}{\textbf{P@10}} & \multicolumn{2}{c|}{\textbf{P@20}}&\textbf{W/T/L} \\
\hline
Baseline & $0.1350$ & -- & $0.2940$ & -- & $0.2345$ & -- & --/--/--\\
+ELR & ${0.1390}$ & $ +2.96\%  $  & \bm{${0.2980}$} & $ +1.36\%  $ & \bm{${0.2375}$} & $ +1.28\%  $ & 46/19/35\\
+TransE & \bm{${0.1392}$} & $ +3.11\%  $& ${0.2950}$ & $ +0.34\%  $ & ${0.2365}$ & $ +0.85\%  $ & 44/16/40\\
\hline
\multirow{2}{*}{} & \multicolumn{7}{c}{ \textbf{QALD-2}}\\
\cline{2-8}
& \multicolumn{2}{c|}{\textbf{MAP}} & \multicolumn{2}{c|}{\textbf{P@10}} & \multicolumn{2}{c|}{\textbf{P@20}}&\textbf{W/T/L}\\
\hline
Baseline& $0.2229$ & --& $0.1529$ & -- & $0.1257$ & -- & --/--/--\\
+ELR & ${0.2197}$ & $ -1.44\%  $& \bm{${0.1671}^{\dagger }$} & $ +9.29\%  $ & ${0.1286}$ & $ +2.31\%  $ & 39/75/26\\
+TransE & \bm{${0.2278}$} & $ +2.20\%  $& ${0.1629}$ & $ +6.54\%  $ & \bm{${0.1321}^{\dagger }$} & $ +5.09\%  $ & 44/68/28\\
\hline
\multirow{2}{*}{} & \multicolumn{7}{c}{ \textbf{ALL}}\\
\cline{2-8}
& \multicolumn{2}{c|}{\textbf{MAP}} & \multicolumn{2}{c|}{\textbf{P@10}} & \multicolumn{2}{c}{\textbf{P@20}}&\textbf{W/T/L}\\
\hline
Baseline & $0.1925$ & -- & $0.2245$ & -- & $0.1798$ & -- & --/--/--\\
+ELR & ${0.2005}^{\dagger }$ & $ +4.16\%  $  & ${0.2307}^{\dagger }$ & $ +2.76\%  $ & ${0.1816}$ & $ +1.00\%  $ & 200/153/132\\
+TransE & \bm{${0.2054}^{\dagger }$} & $ +6.70\%  $& \bm{${0.2346}^{\dagger }$} & $ +4.50\%  $ & \bm{${0.1881}^{\dagger \ddagger }$} & $ +4.62\%  $ & 249/129/107\\
\hline
\end{tabular}}

\end{table}

\subsection{Overall Performance}
In this part, we conduct the overall performance of three models with Coordinate Ascent and RankSVM, as shown in Table~\ref{table:ca} and Table~\ref{table:ranksvm} respectively. 

The entity mention based exact match feature is introduced by Entity Linking incorporated Retrieval model (ELR)~\cite{hasibi2016exploiting} and shows its effectiveness by improving the baseline with Coordinate Ascent and RankSVM almost 1\% and 4\% for the whole data. Then \texttt{+ELR} model shows a significant improvement on the ListSearch test scenario. The ListSearch subtask aims to find related entities which share the same type. And \texttt{+ELR} demonstrates that leveraging entity mention based exact match feature can help to enhance the entity retrieval performance.

The entity embedding is a kind of entity semantic representations and models entity relations in the whole knowledge graph. \texttt{+TransE} model overall improves the Coordinate Ascent based baseline and the RankSVM based baseline significantly by over over 6\% and 5\% respectively. \texttt{+TransE} also illustrates its effectiveness with the significant improvement on SemSearch, ListSearch and QALD-2 test scenarios. And the improvement demonstrates the entity semantic match plays an important role in the task which needs more semantic or linguistic understanding. Both \texttt{+TransE} and \texttt{+ELR} improves baseline model with a small margin on the INDEX-LD test scenario, which illustrates that INDEX-LD only needs keyword matches and the multi-field based entity representation can do well on this scenario. The \texttt{+TransE} model also shows its effectiveness by a large margin improvement with \texttt{+ELR} model, especially on the SemSearch scenario, which illustrates entity embeddings can help model better understand the semantic information of entities.

Overall experiments present the entity effectiveness by a significant improvement especially on the scenarios which need more semantic understanding. Nevertheless, the role of entity based matches in entity retrieval is not clear. Therefore, we explore the importance of the entity based match in the following experiments.

\begin{figure}[t]

    \subfigure[Weight ratio with Coordinate Ascent.] { \label{fig:ca} 
    \includegraphics[width=0.49\linewidth]{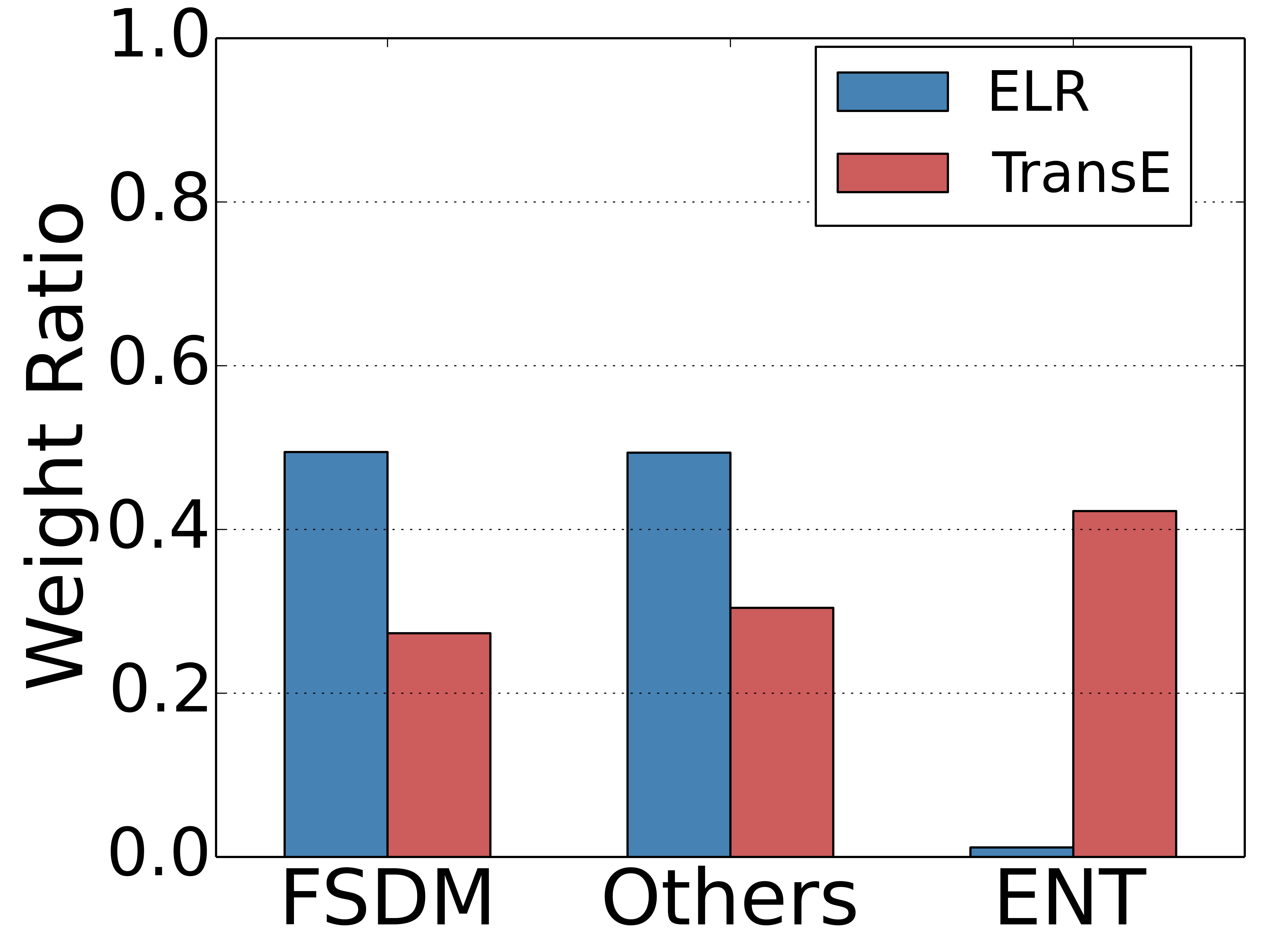}}
    \subfigure[Weight ratio with RankSVM.] { \label{fig:svm} 
    \includegraphics[width=0.49\linewidth]{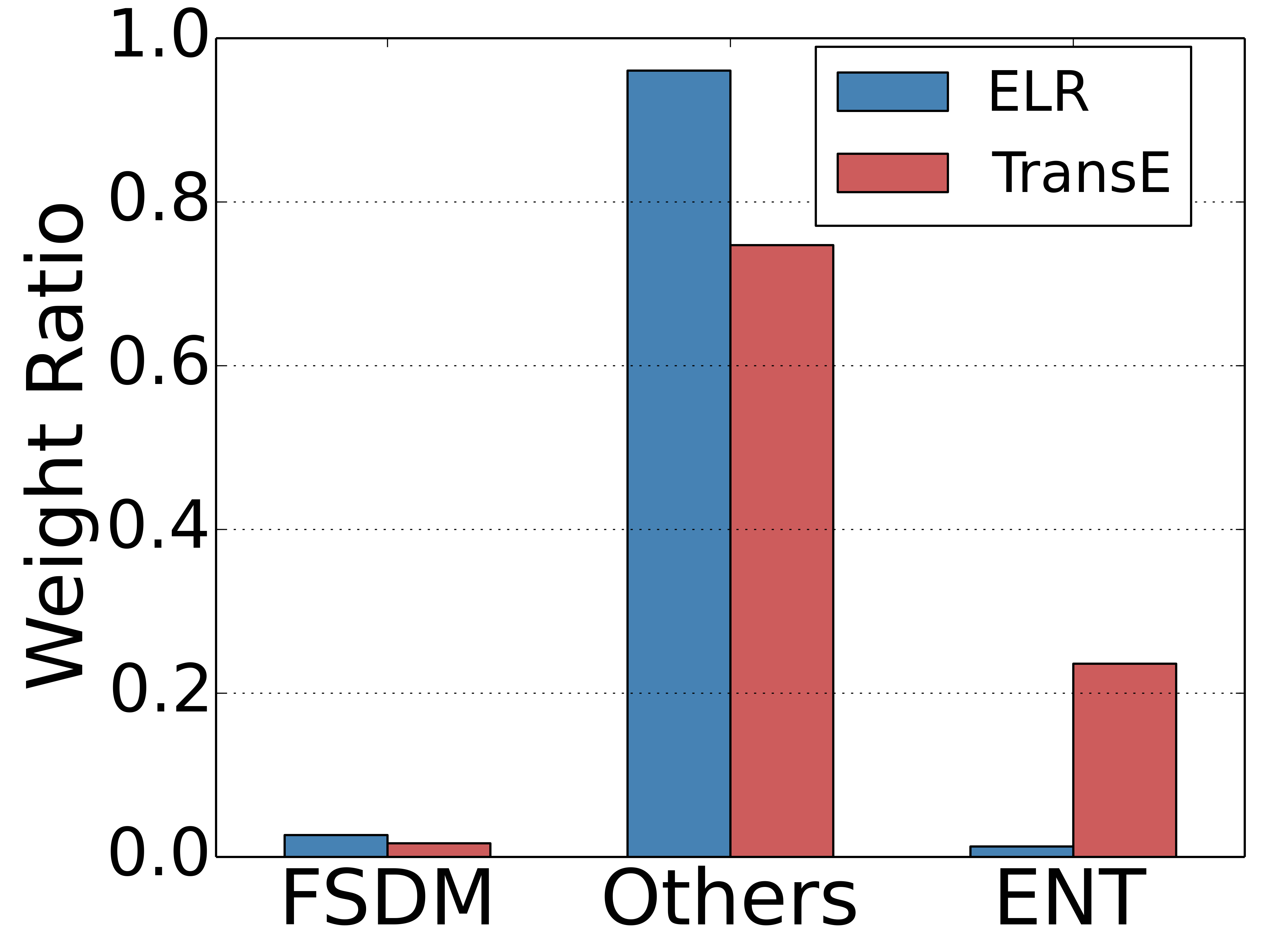}}
    \caption{Feature weight distribution for FSDM based ranking feature, entity based ranking feature (ENT) and other traditional retrieval model based ranking features (Others).}
    \label{fig:weight}
\end{figure}
\subsection{Feature Weight Distribution}
This part presents the weight distribution of ranking features, which are divided into three groups: the FSDM based ranking feature (FSDM), the entity based ranking feature (ENT) and other traditional retrieval model based ranking features (Others), as shown in Figure~\ref{fig:weight}. Then we calculate the percentage of weight given to each type of features by summing up the absolute weight values according to the feature group. For all 27 features (baseline 26 match features with entity based match features), TransE and FSDM features play the most important roles in our model and other traditional ranking features also show their effectiveness especially for RankSVM. On the other hand, TransE shares more weight than the FSDM based ranking feature, confirming the entity semantic match feature is so important for entity retrieval. Moreover, the different weight distributions between ELR and TransE based entity retrieval models demonstrate the entity embedding based semantic match is more effective and important than the entity mention based entity exact match. The semantic information from knowledge graph is brought to the entity retrieval system through the entity embedding and helps entity retrieval models achieve further improvement.

\section{Conclusion}
This paper explores entity embedding effectiveness in entity retrieval with two previous state-of-the-art learning to rank methods which incorporate diverse features extracted from different models and different fields. Entity embedding shows its effectiveness by incorporating entity semantic information from the knowledge graph, which can better model the interaction between query and candidate entities from entity based matches. Experiments on an entity-oriented test collection reveal the power of entity embeddings, especially for the task which needs more semantic and linguistic understanding. Our further analysis reveals that entity embedding based semantic match features plays the same important role as FSDM in entity retrieval and better models query and entity relations than the entity mention based exact match feature. We hope our experiments and models can provide a potential way to better represent entity and leverage semantic information from knowledge graph for entity retrieval systems.

\section*{Acknowledgments}
This work is supported by National Natural Science Foundation of China (NSFC) grant 61532001.
\newpage
\bibliographystyle{splncs04}
\bibliography{mybibliography}

\begin{thebibliography}{10}
\providecommand{\url}[1]{\texttt{#1}}
\providecommand{\urlprefix}{URL }
\providecommand{\doi}[1]{https://doi.org/#1}

\bibitem{balog2006formal}
Balog, K., Azzopardi, L., De~Rijke, M.: Formal models for expert finding in
  enterprise corpora. In: Proceedings of the 29th annual international ACM
  SIGIR conference on Research and development in information retrieval (2006)

\bibitem{balog2011query}
Balog, K., Bron, M., De~Rijke, M.: Query modeling for entity search based on
  terms, categories, and examples. ACM Transactions on Information Systems
  (2011)

\bibitem{balog2013test}
Balog, K., Neumayer, R.: A test collection for entity search in dbpedia. In:
  Proceedings of the 36th international ACM SIGIR conference on Research and
  development in information retrieval (2013)

\bibitem{bendersky2010learning}
Bendersky, M., Metzler, D., Croft, W.B.: Learning concept importance using a
  weighted dependence model. In: Proceedings of the third ACM international
  conference on Web search and data mining (2010)

\bibitem{bordes2013translating}
Bordes, A., Usunier, N., Garcia-Duran, A., Weston, J., Yakhnenko, O.:
  Translating embeddings for modeling multi-relational data. In: Advances in
  neural information processing systems (2013)

\bibitem{chen2016empirical}
Chen, J., Xiong, C., Callan, J.: An empirical study of learning to rank for
  entity search. In: Proceedings of the 39th International ACM SIGIR conference
  on Research and Development in Information Retrieval (2016)

\bibitem{de2007overview}
De~Vries, A.P., Vercoustre, A.M., Thom, J.A., Craswell, N., Lalmas, M.:
  Overview of the inex 2007 entity ranking track. In: International Workshop of
  the Initiative for the Evaluation of XML Retrieval (2007)

\bibitem{demartini2009overview}
Demartini, G., Iofciu, T., De~Vries, A.P.: Overview of the inex 2009 entity
  ranking track. In: International Workshop of the Initiative for the
  Evaluation of XML Retrieval (2009)

\bibitem{hasibi2016exploiting}
Hasibi, F., Balog, K., Bratsberg, S.E.: Exploiting entity linking in queries
  for entity retrieval. In: Proceedings of the 2016 ACM International
  Conference on the Theory of Information Retrieval (2016)

\bibitem{kim2009probabilistic}
Kim, J., Xue, X., Croft, W.B.: A probabilistic retrieval model for
  semistructured data. In: European conference on information retrieval (2009)

\bibitem{liu2018entity}
Liu, Z., Xiong, C., Sun, M., Liu, Z.: Entity-duet neural ranking: Understanding
  the role of knowledge graph semantics in neural information retrieval. In:
  Proceedings of the 56th Annual Meeting of the Association for Computational
  Linguistics (2018)

\bibitem{metzler2005markov}
Metzler, D., Croft, W.B.: A markov random field model for term dependencies.
  In: Proceedings of the 28th annual international ACM SIGIR conference on
  Research and development in information retrieval (2005)

\bibitem{ogilvie2003combining}
Ogilvie, P., Callan, J.: Combining document representations for known-item
  search. In: Proceedings of the 26th annual international ACM SIGIR conference
  on Research and development in informaion retrieval (2003)

\bibitem{pound2010ad}
Pound, J., Mika, P., Zaragoza, H.: Ad-hoc object retrieval in the web of data.
  In: Proceedings of the 19th international conference on World wide web. ACM
  (2010)

\bibitem{robertson2004simple}
Robertson, S., Zaragoza, H., Taylor, M.: Simple bm25 extension to multiple
  weighted fields. In: Proceedings of the thirteenth ACM international
  conference on Information and knowledge management. pp. 42--49. ACM (2004)

\bibitem{wang2012overview}
Wang, Q., Kamps, J., Camps, G.R., Marx, M., Schuth, A., Theobald, M., Gurajada,
  S., Mishra, A.: Overview of the inex 2012 linked data track. In: CLEF (2012)

\bibitem{zhiltsov2015fielded}
Zhiltsov, N., Kotov, A., Nikolaev, F.: Fielded sequential dependence model for
  ad-hoc entity retrieval in the web of data. In: Proceedings of the 38th
  International ACM SIGIR Conference on Research and Development in Information
  Retrieval (2015)

\end{thebibliography}

\end{document}